\documentclass[prl,twocolumn,superscriptaddress,letterpaper]{revtex4}
\usepackage{graphicx,amssymb,bm,natbib,amsfonts,amsmath}
\usepackage{hyperref}

\begin{document}

\title{Quasi-Freestanding Multilayer Graphene Films on the Carbon Face of SiC}

\author{D. A. Siegel}
\affiliation{Department of Physics, University of California,
Berkeley, CA 94720, USA}
\affiliation{Materials Sciences Division,
Lawrence Berkeley National Laboratory, Berkeley, CA 94720, USA}

\author{C. G. Hwang}
\affiliation{Materials Sciences Division,
Lawrence Berkeley National Laboratory, Berkeley, CA 94720, USA}

\author{A. V. Fedorov}
\affiliation{Advanced Light Source, Lawrence Berkeley National Laboratory, Berkeley, CA 94720, USA}

\author{A. Lanzara}
\affiliation{Department of Physics, University of California,
Berkeley, CA 94720, USA}
\affiliation{Materials Sciences Division,
Lawrence Berkeley National Laboratory, Berkeley, CA 94720, USA}

\date{\today}

\begin{abstract}
The electronic band structure of as-grown and doped graphene grown on the carbon face of SiC is studied by high-resolution angle-resolved photoemission spectroscopy, where we observe both rotations between adjacent layers and AB-stacking. The band structure of quasi-freestanding AB-bilayers is directly compared with bilayer graphene grown on the Si-face of SiC to study the impact of the substrate on the electronic properties of epitaxial graphene. Our results show that the C-face films are nearly freestanding from an electronic point of view, due to the rotations between graphene layers.
\end{abstract}

\maketitle
One of the most substantial problems that the graphene community is faced with is a choice of substrate that preserves the unique properties of the Dirac charge carriers in graphene.  Most substrates require tradeoffs between ease of large-scale sample growth and the strength of the substrate interaction.  One of the earliest forms of graphene was grown on transition metal substrates like nickel \cite{Karu,Land}, where the growth process itself is straightforward but the interaction with the substrate is very strong \cite{Rosei, Shikin}.  More recently, free-standing graphene has been isolated through mechanical exfoliation \cite{NovoselovScience}, but the process is time-consuming and unreliable.  The ideal graphene system for the purposes of both scientific studies and industrial applications would be one where large-scale sample growth is simple and efficient, and the graphene is relatively free-standing.  Graphene grown on the carbon face of SiC, SiC($000\overline{1}$) \cite{BergerScience} might be one such system.  Despite the large number of graphene layers, the different orientations between adjacent layers (most commonly $\pm2^{\circ}$ and 30$^{\circ}$ rotation with respect to the substrate) \cite{forbeauxcface,HassPRB,HassReview} causes them to decouple from one another, resulting in a system whose transport and electronic properties closely match those of freestanding monolayer graphene \cite{BergerScience,Orlita,dosSantos,SadowskiPRL,Sprinkle}. 
On the other hand, recent magnetospectroscopy and localization measurements suggest a more complicated role of the interlayer interaction \cite{XWu} and the possible presence of AB-stacked (Bernal or rhombohedral) domains \cite{SadowskiSSC}.  In addition, x-ray measurements of C-face graphene indicate an average lattice spacing that lies between that of AB-stacked graphene and of fully rotationally disordered films \cite{HassPRB}.  This is completely different from the case of graphene grown on the Si-face of SiC, where all adjactent layers are coupled due to the AB-stacking \cite{PRWallace, OhtaZdep, OhtaBilayer}.

To shed light on the role of the interlayer interaction and to fully characterize the electronic structure of these samples, it is of fundamental importance to directly measure their band structure and in particular to study the $\pi$ bands near the Dirac point.  Angle-resolved photoemission spectroscopy (ARPES) is the ideal tool to directly measure the electronic band structure of graphene and determine the number of AB-stacked layers by measuring the number of $\pi$ bands.  The thickness of AB-stacked films can be verified by measuring the variation in photoemission intensity of these bands along the $k_z$ (out-of-plane) direction \cite{OhtaZdep,ShuyunPhysicaE}.

In this Rapid Communication we present high-resolution ARPES studies of the $\pi$ bands of epitaxial graphene grown on the carbon face of SiC. On every sample, we observe the decoupled bandstructure of freestanding monolayer graphene, AB-stacked bilayer graphene, and other AB-stacked few-layer-graphene bandstructures, although the relative amounts of AB-stacking can vary from sample to sample.
This indicates the prevalence of both AB-stacking and the decoupling of adjacent layers by azimuthal rotations to create freestanding few-layer AB-stacked systems.  We discuss how to reconcile these findings with transport properties which appear to be identical to those of single-layer graphene.  Finally, we show how the bandstructure of freestanding bilayer graphene grown on the carbon face of SiC differ from those of bilayer graphene grown on the silicon face of SiC, in order to demonstrate the impact of the substrate on the electronic properties.

Samples were grown on the C-terminated face of SiC substrate as previously reported \cite{BergerScience,HassReview}.  High-resolution ARPES data were taken at BL12.0.1 of the Advanced Light Source at a temperature of 15$^{\circ}$K after annealing samples to 1300$^{\circ}$K using photon energies from 42eV to 80eV.  The vacuum was better than $3\times10^{-11}$ Torr.

Low energy electron diffraction \cite{forbeauxcface} and x-ray diffraction \cite{HassPRB, HassReview} measurements of the C-face growth of graphene show a distribution of rotational orientations of the graphene planes with respect to the SiC substrate.  Since the sample has rotational domains, ARPES data taken radially outwards from the $\Gamma$-point, along the $\Gamma$-K direction (see cartoon in Fig. 1, top center), are generally sharper than data taken along the azimuthal direction (the K-K' direction), so the ARPES spectra presented here are shown along the $\Gamma$-K orientation.  Moreover, the bilayer band intensity along the K-K' direction is greatly suppressed by photoemission matrix elements (even more so than monolayer graphene).

\begin{figure}
\includegraphics[width=8.5 cm] {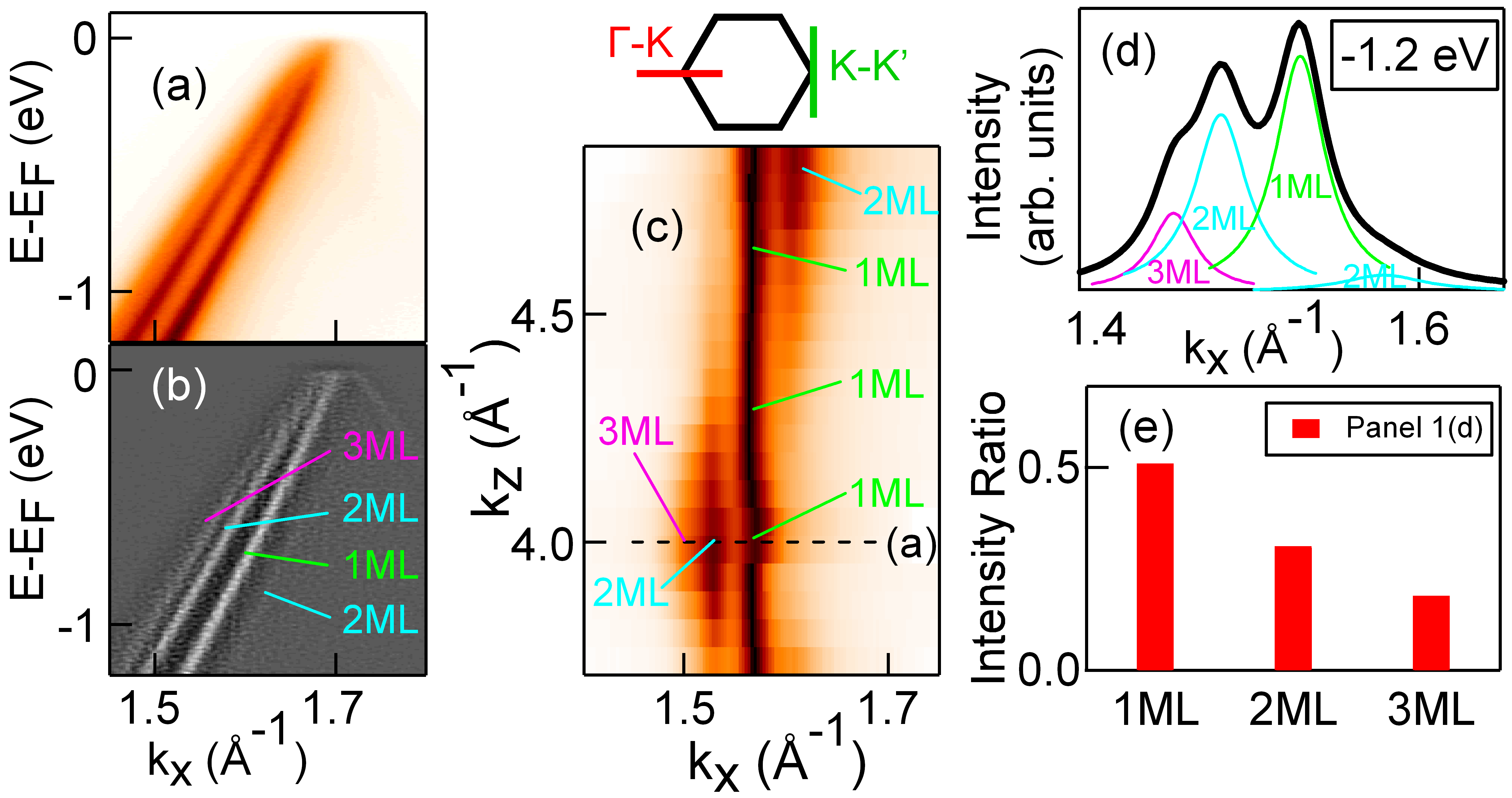}
\caption{(Color online) (a) ARPES dispersions taken along the $\Gamma$-K direction at k$_{z}$ = 4.0 \AA$^{-1}$, showing bands that correspond to a superposition of signals from AB-stacked films of different thicknesses.  (b) The second derivative of panel (a) enhances the weaker multilayer sidebands. (c) Photon energy or k$_{z}$-dependence of photoemission intensity, at 1 eV binding energy, shows that the bands of panel (a) correspond to different thicknesses of AB-stacked films.  (d) Four Lorentzian peaks were fit to ten MDCs to obtain the data in (e).  (e) A comparison of the intensity ratios of the bands in (a) to the intensity ratios one should expect from previously measured x-ray data.}
\end{figure}

Figure 1(a) shows raw ARPES data that is typical of our samples.  It was taken at a K-point (rotated by 2$^{\circ}$ from the SiC lattice) along the $\Gamma$-K direction, and there are several bands that disperse towards the Dirac point, which is near the Fermi level.  This is surprising because the Dirac cone of monolayer graphene has only a single band, whereas AB-stacked graphite has only two.  The extra bands are not due to rotated Dirac points, which has been ruled out by an examination of the Fermi surface and other constant-energy maps.  The band velocities and Dirac point momenta also disagree with the predicted bandstructure of rotational supercells\cite{Latil}.  Instead, the bands in Fig. 1 correspond to a superposition of the valence band dispersions of single-layer, double-layer, and multi-layer AB-stacked graphene \cite{partoens,mccann1,mccann2}.  
For thin AB-stacked graphene films, the thickness can be determined from photoemission by measuring the number of $\pi$ bands, and can be confirmed by observing the intensity oscillations along the $k_z$ direction \cite{OhtaZdep,ShuyunPhysicaE}.  In an AB-stacked arrangement, the number of $\pi$ bands is equal to the number of graphene layers.  The relative intensities of these ARPES bands vary, and occasionally vanish, as the ARPES signal is scanned along the surface of the sample.

From the separation of the momentum distribution curves (MDCs) at $E_F$ (not shown) we conclude that the sample is slightly p-doped: for monolayer graphene the doping is  8$\times$10$^{10}$ cm$^{-2}$ \cite{Sprinkle,MonolayerPaper}), as expected for a freestanding graphene sample \cite{NovoselovScience, PKim, Andrei}.  This differs significantly from graphene grown on the Si face of SiC, where monolayer graphene is typically n-doped by $\sim$1$\times$10$^{13}$ cm$^{-2}$ \cite{OhtaZdep}.  In graphene grown on the Si-face of SiC, a Schottky potential forms between the graphene bilayer and the adjacent SiC substrate, which induces a charge transfer onto the graphene sheet\cite{SeyllerSchottky,Sebastien}, resulting in a higher doping level \cite{ZhouGap,OhtaBilayer}.  In contrast, graphene grown on the C-face of SiC can be many layers thick \cite{HassReview}, where the topmost films are well-separated from the SiC substrate, sitting instead on a thick graphitic substrate and electronically decoupled by rotational faults.  This results in the lower doping level and lower overall interaction with the substrate in graphene grown on the C-face of SiC.  Illustrations of the real-space structures are given in Fig. 2.

\begin{figure} \includegraphics[width=8.5cm]{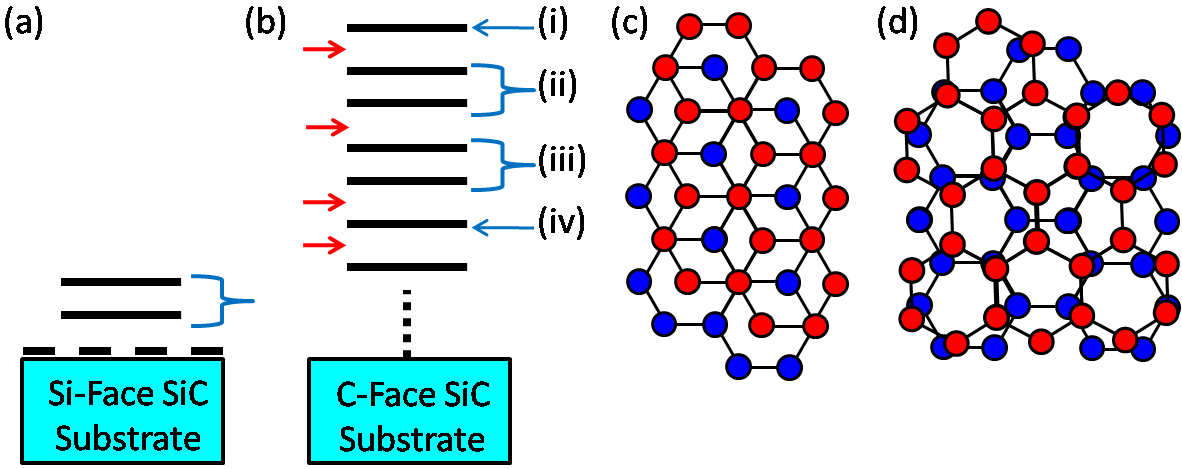}
\caption{(Color online) (a) A cartoon illustrating bilayer graphene grown on the Si-face of SiC, including the SiC substrate, carbon-rich buffer layer (horizontal dotted line) and bilayer graphene planes (horizontal solid lines).  (b) A cartoon illustrating the stackings in C-face graphene.  Arrows mark rotational faults.  Layer groups (i)-(iv) correspond to 1ML, 2ML, 2ML, 1ML AB-stacked domains, respectively.  (c) An AB-stacked bilayer pair.  (d) A rotated bilayer pair.}\end{figure}

A standard method of graphene thickness characterization by ARPES is the measurement of the photon energy, or k$_{z}$, intensity-dependence of the graphene valence bands \cite{OhtaZdep,ShuyunPhysicaE}.  This dispersion is shown for the C-face sample in panel (c).  The most intense band is the vertical straight line that corresponds to monolayer graphene; in rare cases, only this single line is present.  In addition to this straight line, one can discern an alternating double line due to the split bands of bilayer graphene, and a shoulder near k$_{z}$ = 4.0 \AA$^{-1}$ corresponds to bands from thicker films.  The periodicity of the bilayer band intensity matches that reported by Ohta et al. \cite{slonczewski,OhtaZdep} for AB-stacking on Si-face samples.  This indicates that the AB-stacking in C-face and Si-face samples have the same out-of-plane lattice constant, to within 0.8\%.  These intensity modulations, typical of every measured C-face sample (including one UHV-graphetized 6H-SiC sample and several furnace-graphetized 4H-SiC samples), confirm the presence of AB-stacked graphene domains.

Since a single ARPES beamspot probes many domains on the surface of the sample (which can also result in multiple Dirac cones in a single ARPES spectrum), a statistical distribution of film thicknesses can be obtained from a single ARPES image \cite{OhtaLEEM}.  This can be done by Lorentzian fitting to the MDCs extracted from the dispersion in panel (a) (a typical raw MDC with its fit is shown in panel (d)).  The high spectral weight from AB-stacked ARPES bands, shown in panel (e), are typical of all measured samples.  The fact that much of the sample forms AB-stacked structures is surprising in light of transport experiments where bulk samples behave like monolayer graphene. 

The reason why many transport experiments do not show signatures of multilayer domains might be related with: a) the smaller Fermi velocity of multilayer graphene; b) the possible presence of a bandgap in undoped bilayer and few-layer graphene \cite{OhtaBilayer,ZhouGap}; c) the local nature of some experiments, where a single monolayer domain may certainly be probed \cite{MillerScience}; d) the bulk of these samples may have a lower AB-density than the surface probed by ARPES; e) overall variations in the AB-stacking probability from sample to sample; and perhaps most importantly, f) interactions across rotational faults.  Coupling between rotated planes may result in supercell bandstructures with linear dispersions that act like monolayer graphene \cite{Latil}.  However, we could not obtain any evidence of these supercell bands, perhaps due to experimental limitations.

\begin{figure} \includegraphics[width=8.5cm]{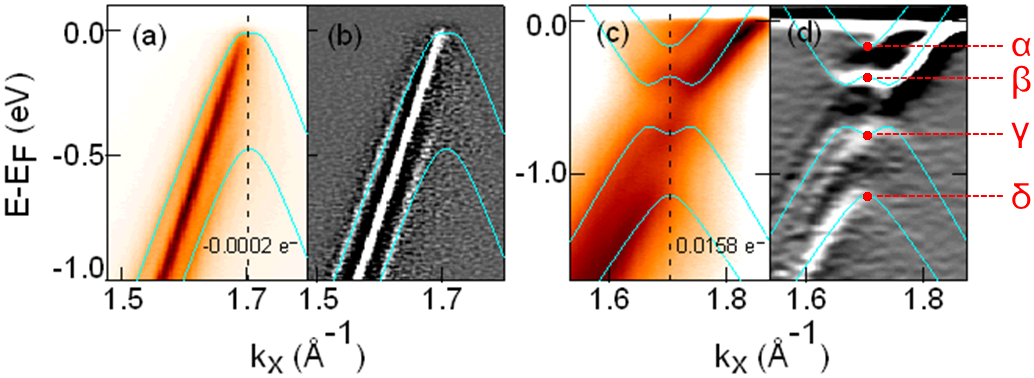}
\caption{(Color online) (a) ARPES dispersions taken along the $\Gamma$-K direction at k$_{z}$ = 4.5 \AA$^{-1}$ for the as-grown sample.  Tight-binding dispersions have been added for the bilayer bands as a guide to the eye.  The vertical dotted line in panels (a) and (c) give the position of the K-point. (b) is a second derivative of (a).  (c) ARPES dispersions taken along the $\Gamma$-K direction at k$_{z}$ = 4.5 \AA$^{-1}$, electron-doped by 0.0158 electrons per unit cell by potassium deposition.  Tight binding dispersions have been added as a guide to the eye.  (d) is a second derivative of (c).  The K point in panel (d) have been marked for each of the bilayer bands.}\end{figure}

To address the way the stacking takes place in these multilayer domains, we focus on bilayer graphene and use the tight binding model, as in a previous study \cite{OhtaBilayer,partoens,mccann1,mccann2}, to model the onsite Coulomb potential $U$ and the out-of-plane nearest neighbor interaction, $\gamma_1$.
To obtain $U$ and $\gamma_1$, we electron-dope the sample by potassium deposition and extract the bilayer graphene band positions at the K-point for each doping.  According to the tight binding, the band positions at the K point are given by\\
\\
$\varepsilon_{\sigma,\tau}(k=K)=(-1)^{\sigma}[ \frac{(1+(-1)^{\tau})}{2}\gamma_1^2+\frac{U^2}{4} ]^{1/2}$,\\
\\
where $\sigma,\tau=\pm1$ correspond to the choice of band.

In Figs. 3(a) and 3(b) we show data for the as grown sample at k$_{z}$=4.5 \AA$^{-1}$.  At this value of k$_{z}$, the two bilayer bands have equal intensity, although the monolayer band is more intense than both.  The hatlike structure that is often associated with bilayer graphene \cite{ZhouPRL,Nilsson} is absent for the undoped bands, which implies that the potential difference between the two bilayer graphene planes is small, a consequence of the distance from the SiC substrate.  The dispersions for the doped sample are shown in panels (c) and (d), where the potential created by the adsorbed potassium atoms give the bilayer bands a hatlike shape.

\begin{figure} \includegraphics[width=8.5cm]{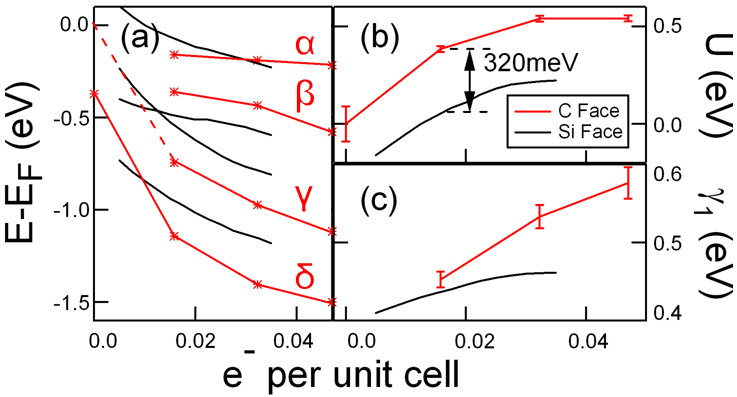}
\caption{(Color online) (a) Comparison of band energies at the K-point (in Fig. 2 these are the energies where bands cross the vertical dotted lines, as marked by Greek letters in panel 2d) for bilayer graphene grown on the C-face in red (grey) and Si-face (black) of SiC, as a function of doping.  Si-face data was obtained from Ref. \cite{OhtaBilayer}. (b) Comparison of $U$ as a function of doping. The error bar for the leftmost data point gives an upper limit from a tight-binding fit.  (c) Comparison of $\gamma_1$ as a function of doping.}\end{figure}

In Fig. 4 we summarize the results of the fitting to several dopings. These are compared with data from graphene grown on the Si face of SiC \cite{OhtaBilayer}, where the interaction with the substrate is strong \cite{ZhouGap}, to study how the tight binding parameters are affected by an increase in the interlayer interaction.  Panel (a) shows the band positions at the K point of both types of sample as a function of doping, (b) shows $U$, and (c) shows $\gamma_1$.  We find that all of these parameters have a larger value for the C-face graphene than for the Si-face graphene.

For $U$, this is a demonstration of the freestanding nature of the C-face samples.  
For the Si-face samples the graphene layer is separated only by a carbon rich "`buffer layer"' from the substrate and is hence subject to a strong interaction with the substrate. This shifts the Dirac point of monolayer graphene by $\sim$400 meV \cite{OhtaZdep,ZhouGap}, and results in a large potential difference between the as-grown sheets of bilayer graphene.  On the contrary, for the C-face sample the graphene layer is separated from the SiC substrate by a thick graphitic film, resulting in an almost negligible interaction with the substrate.  This induces only a small shift of the Dirac point of monolayer graphene by only 39 meV from the Fermi level (smaller by more than an order of magnitude; the charge transfer is smaller by two orders of magnitude), and the undoped bilayer graphene is relatively free-standing with little potential difference between the as-grown sheets.  As a result, the value of $U$ is offset by $\sim$320meV between graphene grown on the Si-face and the more freestanding C-face.  This is outlined schematically in Fig. 5, where it is shown how the built-in field of the as-grown Si-face sample adds an overall offset to the doping-dependence.

\begin{figure} \includegraphics[width=8.5cm]{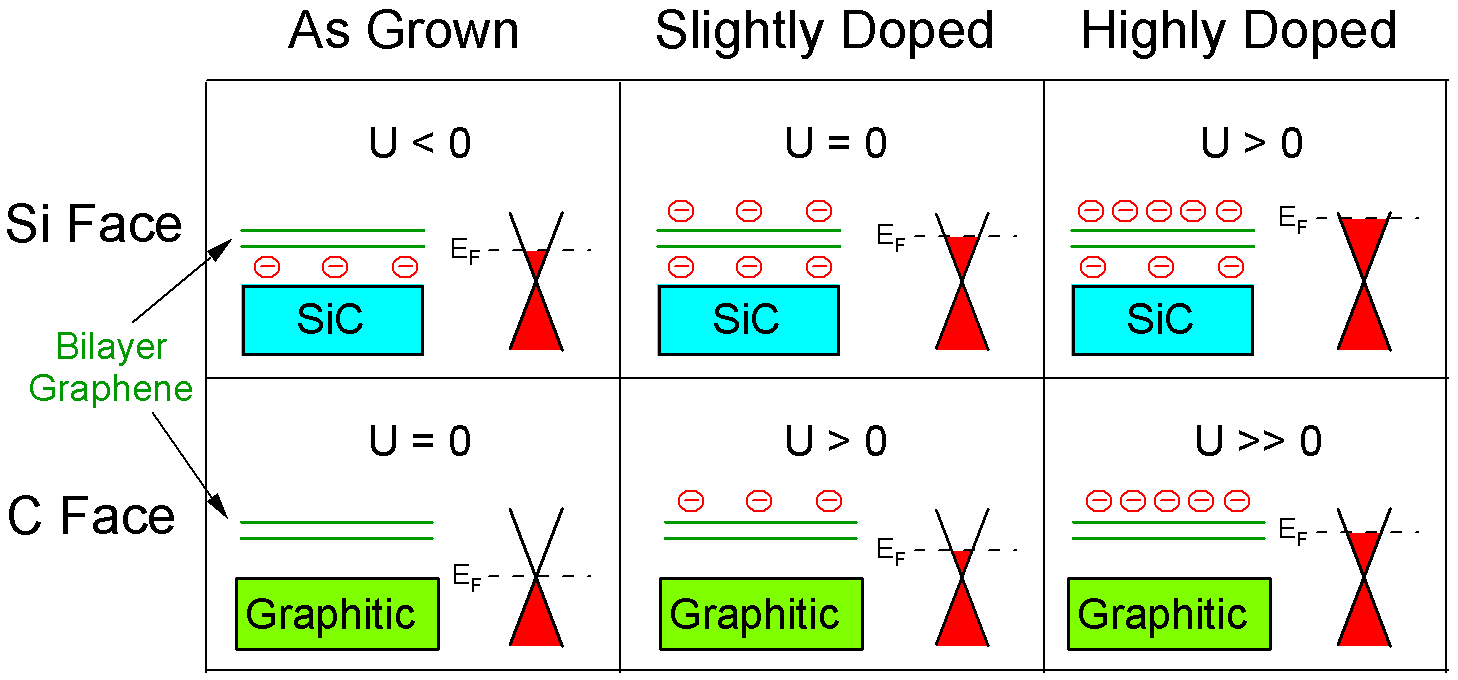}
\caption{(Color online) Illustrations of the interlayer potential difference $U$ for bilayer graphene on the carbon and silicon faces of SiC for several dopings.  On the Si face, the as-grown junction between bilayer graphene and the SiC substrate forms a Schottky potential and a buildup of charge near the interface, which results in a finite $U$.  On the C face, azimuthal rotations between bilayer graphene and the graphitic substrate result in a much smaller substrate interaction and a smaller $U$.  Electron-doping the sample with potassium increases $U$ as indicated.}\end{figure}

The larger values of $\gamma_1$ reflect a greater overlap between nearest neighbor out-of-plane orbitals for the C-face sample.  For the Si-face samples, changes were attributed to a shorter screening length due to the increased charge density \cite{OhtaBilayer}.  The same is true when changing substrates; the graphitic film on the surface of the C-face sample has a higher dielectric constant than that of SiC, which could help screen the interlayer interaction and result in a slightly smaller lattice spacing.  This might be responsible for the overall offset in $\gamma_1$.  The presence of the carbon-rich buffer layer on the Si-face of SiC \cite{Forbeaux,Seyller} could also be responsible for a change in $\gamma_1$.

In conclusion, we have demonstrated that the electronic properties of the graphitized carbon face of 6H-SiC are a good example of quasi-freestanding single \textit{and} multi-layer AB-stacked graphene, as demonstrated by the measured conical dispersion characteristic of free graphene sheets in the presence of higher dielectric screening.  In particular, we demonstrated the intrinsic properties of bilayer graphene by extracting some of the tight-binding parameters and comparing them to those of samples with a stronger substrate interaction.  The presence of multilayer graphene on the carbon-face samples may enable future studies and growth of free-standing multilayer films and a better understanding of the effects of a substrate on the transport properties of graphene films.

\begin{acknowledgments}
We would like to thank Walt de Heer, Claire Berger, and Yike Hu for providing us with the high quality graphene samples studied in this paper.
Useful discussion with Antonio Castro Neto, Ed Conrad and Mike Sprinkle are also acknowledged.
ARPES work was supported by the Director, Office of Science, Office of Basic Energy Sciences, Materials Sciences and Engineering Division, of the U.S. Department of Energy under Contract No. DE-AC02-05CH11231.  Sample growth was supported by the MRSEC under NSF Grant No. DMR-0820382
\end{acknowledgments}

Correspondence and requests for materials should be addressed to Alanzara@lbl.gov.

\end{document}